\documentclass[aps,prl,twocolumn,superscriptaddress,showpacs]{revtex4}
\usepackage{float}
\usepackage{graphicx}
\usepackage{amssymb}
\begin{document}

\title{Crosslinked networks of stiff filaments exhibit negative normal stress}

\author{Enrico Conti}
\affiliation{Division of Physics and Astronomy, Vrije Universiteit,
1081 HV Amsterdam, The Netherlands}

\author{Fred C. MacKintosh}
\email{fcm@nat.vu.nl} \affiliation{Division of Physics and
Astronomy, Vrije Universiteit, 1081 HV Amsterdam, The Netherlands}

\date{\today}

\begin{abstract}
Motivated by recent experiments showing that stiff biopolymer gels exhibit highly unusual negative normal elastic stresses, we develop a computational model for stiff polymer networks subject to large strains. In all cases, we find that such networks develop normal stresses that are both negative and of magnitude comparable to the corresponding shear stress. We find that these normal stresses coincide with other nonlinearities in our networks, and specifically with compressive bucking of the individual filaments. Our results suggest that negative normal stresses are a characteristic feature of stiff (bio)polymer gels that have been shown to exhibit strong nonlinear elastic properties.
\end{abstract}


\maketitle
Networks of semi-flexible polymers such as those that make up the cytoskeleton of plant and animal cells have been shown to have rich mechanical and rheological properties.
One of the hallmarks of their mechanics is a highly nonlinear elastic response to stress and strain\cite{JanmeyJCB91,Gardel04,Storm05,Wagner06,Kasza,Chaudhuri07}, including dramatic stiffening under shear.
A striking example of such nonlinearity has been the recent demonstration in such systems of highly unusual negative normal stresses---e.g., in which a sample will tend to contract along an axis perpendicular to the direction of shear\cite{Janmey07}.
Normal stresses, in general, are a nonlinear phenomenon, since their sign cannot depend on the direction of shear, for symmetry reasons.
But, most materials tend to expand when sheared, as has been known at least since the classic experiments of Poynting nearly 100 years ago\cite{Poynting1909}, in which he showed that elastic rods extend axially when twisted.
Another familiar example of this is the tendency of granular materials to dilate when sheared, as can be seen by the fact that wet sand tends to dry out around our feet when we walk on the beach. Few materials have been found to develop negative normal stresses.
Liquid crystalline polymers\cite{LC}, nanotubes\cite{Nanotubes} and emulsions\cite{Pasquali} are examples of such systems, which show rather weakly negative normal stresses in a range of applied shear rates.
By contrast, semi-flexible polymer networks exhibit negative normal stresses of comparable magnitude to the observed shear stress, and this is observed in the \emph{elastic} response of such networks.

Here, we simulate networks of stiff rods and show that these networks generally exhibit negative normal stresses (NNS) comparable to or larger than the shear stresses, depending on network structure and the strain applied.
It has previously been shown theoretically that entropic effects can result in such NNS effects\cite{Janmey07}.
We also find such anomalous NNS effects in a purely mechanical (athermal) model of stiff rods.
We study the dependence of both shear and normal stresses on both the density and individual mechanical properties of the constituent filaments.
Although both entropic and enthalpic networks can exhibit NNS, their predicted dependence on network properties such as density are opposite, providing a possible experimental way to distinguish the important roles of entropic and enthalpic effects.

\begin{figure}
\centering
\includegraphics[width=230 pt]{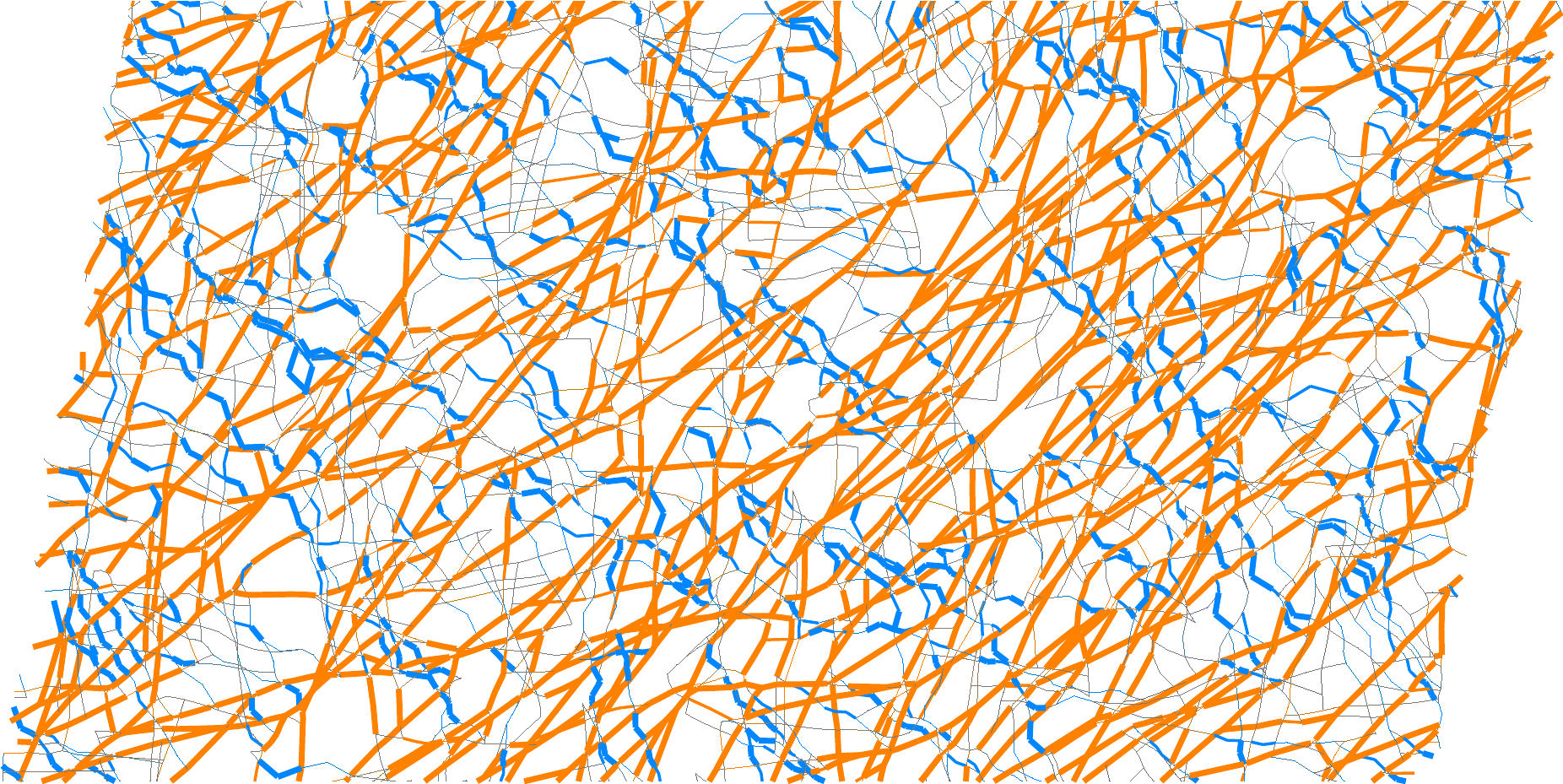}
\caption{A portion of a typical network strained at $\gamma\simeq 0.16$. Red rods (color online) are stretched segments of filaments while blue ones are compressed. Filaments oriented around $45^{\circ}$ are stretched while the ones at about $135^{\circ}$ are compressed and buckled. Horizontal boundaries are rigidly displaced according to a macroscopic strain $\gamma$ while the sides are connected by periodical boundary conditions. } \label{PictureNetwork}
\end{figure}

We construct model networks as follows.
A number of straight filaments of fixed length $L$ and random orientation are deposited on a $W\times W$ square.
Every time two filaments cross each other they are linked together by a free hinge.
Filaments that cross the upper or lower boundary are rigidly attached to the boundary and dangling ends are removed, while periodic boundary conditions are applied to the left and right edges of the square. The deposition stops as soon as the desired concentration is reached.
We characterize this concentration by the ratio $L/\ell_c$, where $\ell_c$ is the average distance between cross-links along a filament.
The network is then refined by adding midpoints between any pair of consecutive cross-links along the same filament.
An energy functional is associated with the position of the modeled points, both cross-links and midpoints, in such a way as to implement a discrete worm like chain approximation for every filament.
For any consecutive pair of points we consider the stretching energy $\delta \mathcal{H}_{\mbox{\tiny{STRETCH}}}$ as a function of the distance $\ell$ and the initial distance $\ell_0$ via a stretching stiffness $\mu$.
\begin{equation}
\delta \mathcal{H}_{\mbox{\tiny{STRETCH}}} = \frac{\mu}{2}\left(\frac{\delta\ell}{\ell_0}\right)^{2}\ell_0\label{stretch}
\end{equation}
The bending energy is evaluated for every pair of consecutive segments along a filament:
\begin{equation}
\delta \mathcal{H}_{\mbox{\tiny{BEND}}} = \frac{\kappa}{2}\left(\frac{\delta\theta}{\ell'}\right)^{2}\ell',\label{bend}
\end{equation}
where $\delta\theta$ is the angular deflection of one segment relative to the other, $\ell'$ is the average length of the two segments and $\kappa$ is the bending stiffness.
The total energy is, then, the sum of these energies over the entire network.
A shear of this network is performed by displacing the upper boundary with respect to the lower one, followed by a minimization of the elastic energy over the unconstrained internal coordinates (cross-links and midpoints) of the network using a nonlinear conjugate gradient technique \cite{NRC}.
Finite deformations are accomplished in a stepwise manner.
The resulting shear (normal) stresses are then determined from the forces parallel (perpendicular) to the displaced boundaries.

We simulated networks of size $W=7$ of filaments with length $L=1$, $2$ and different network densities $L/\ell_c=9$, $11$, $13$, $15$, $20$, $30$, $40$, $50$, $60$.
The bending rigidity $\kappa$ is varied between $10^{-7}$ and $10^{-2}$ keeping a constant $\mu=1$.
Successive boundary deformations are applied in logarithmically-spaced steps in order to explore the strain elastic response at both small and large strains. In Fig.~\ref{PictureNetwork} is depicted a relatively dense network ($L/\ell_c=15$).
Under a purely affine shear, the filaments initially oriented near $45^{\circ}$ to the strain direction are stretched (indicated in red) and aligned, while those oriented near $135^{\circ}$ are compressed (blue) and eventually buckle.
At the filament scale buckling occurs on wavelengths shorter than $L$ due to the connectivity with the surrounding network through the cross-links, in contrast to classical Euler buckling at the fundamental mode\cite{Landau}.

\begin{figure}
\centering
\includegraphics[width=250 pt]{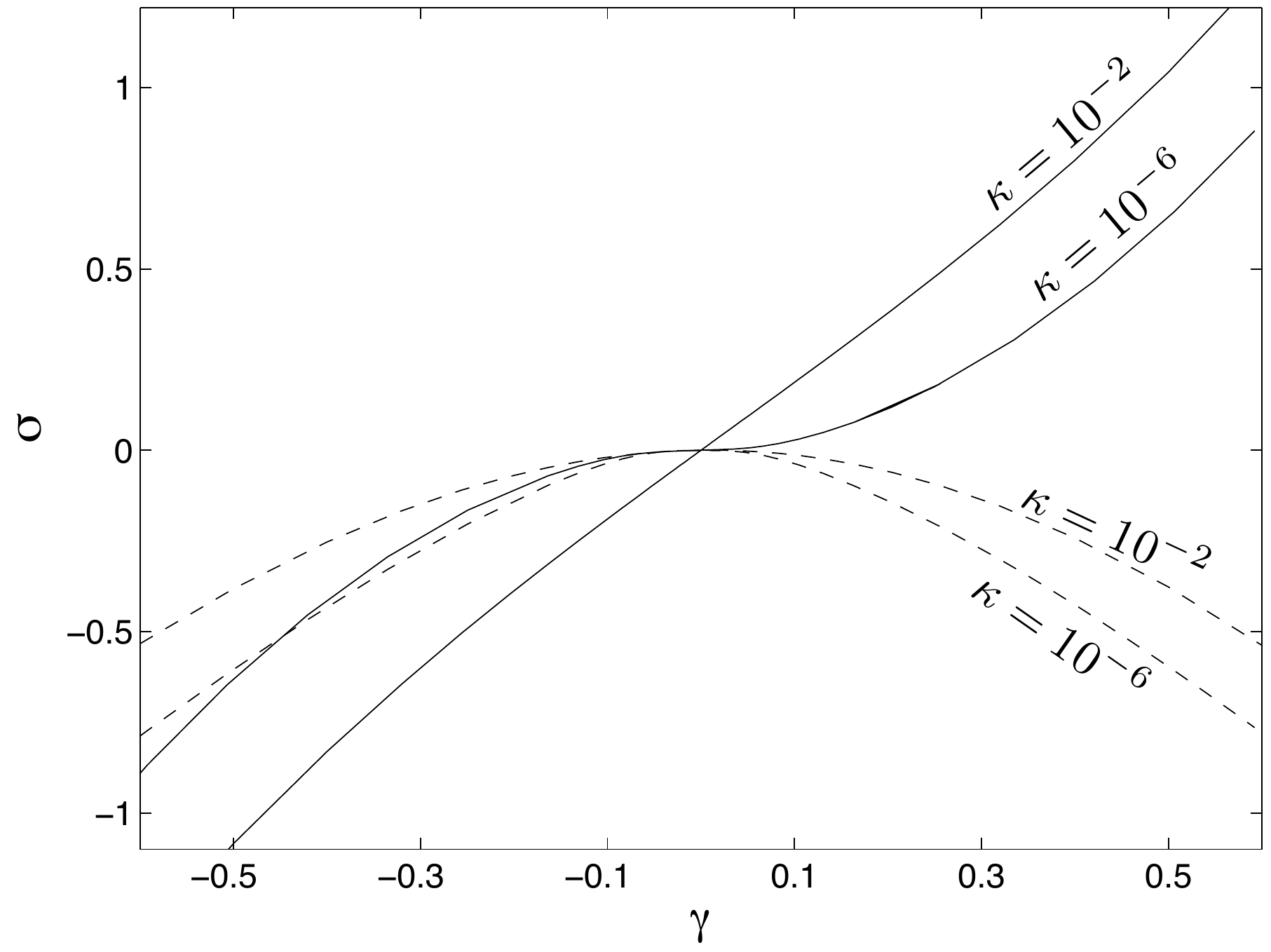}
\caption{Raw shear (solid) and normal (dashed) stresses versus strain for different bending stiffnesses.} \label{SandN}
\end{figure}

In Fig.~\ref{SandN} are shown the normal and shear stresses, $\sigma_N$ and $\sigma_S$, for a single network ($L/\ell_c=13$) in response to an imposed strain $\gamma$ for various bending stiffnesses $\kappa$.
The shear response starts linearly and eventually stiffens weakly, while the normal response is negative (the network wants to contract).
For an ensemble-averaged network with isotropically oriented filaments, the normal response must, by symmetry, be an even function of the shear deformation.
Finite size effects, however, result in a small violation of this for any specific network studied.
The shear stress increases with increasing $\kappa$ while the magnitude of the normal component exhibits the opposite trend.
This behavior is qualitatively consistent with the previous suggestion that the normal stresses arise from the asymmetry of the extensional response of single filaments\cite{Janmey07}, since stiffer filaments are expected to bend less\cite{Head03,Wilhelm03} and the energy in Eq.\ (\ref{stretch}) is symmetric.
Increased bending, and eventual buckling, lead to an increasingly asymmetric extensional response, with softening under compression\cite{FloppyModes}.

\begin{figure}
\centering
\includegraphics[width=250 pt]{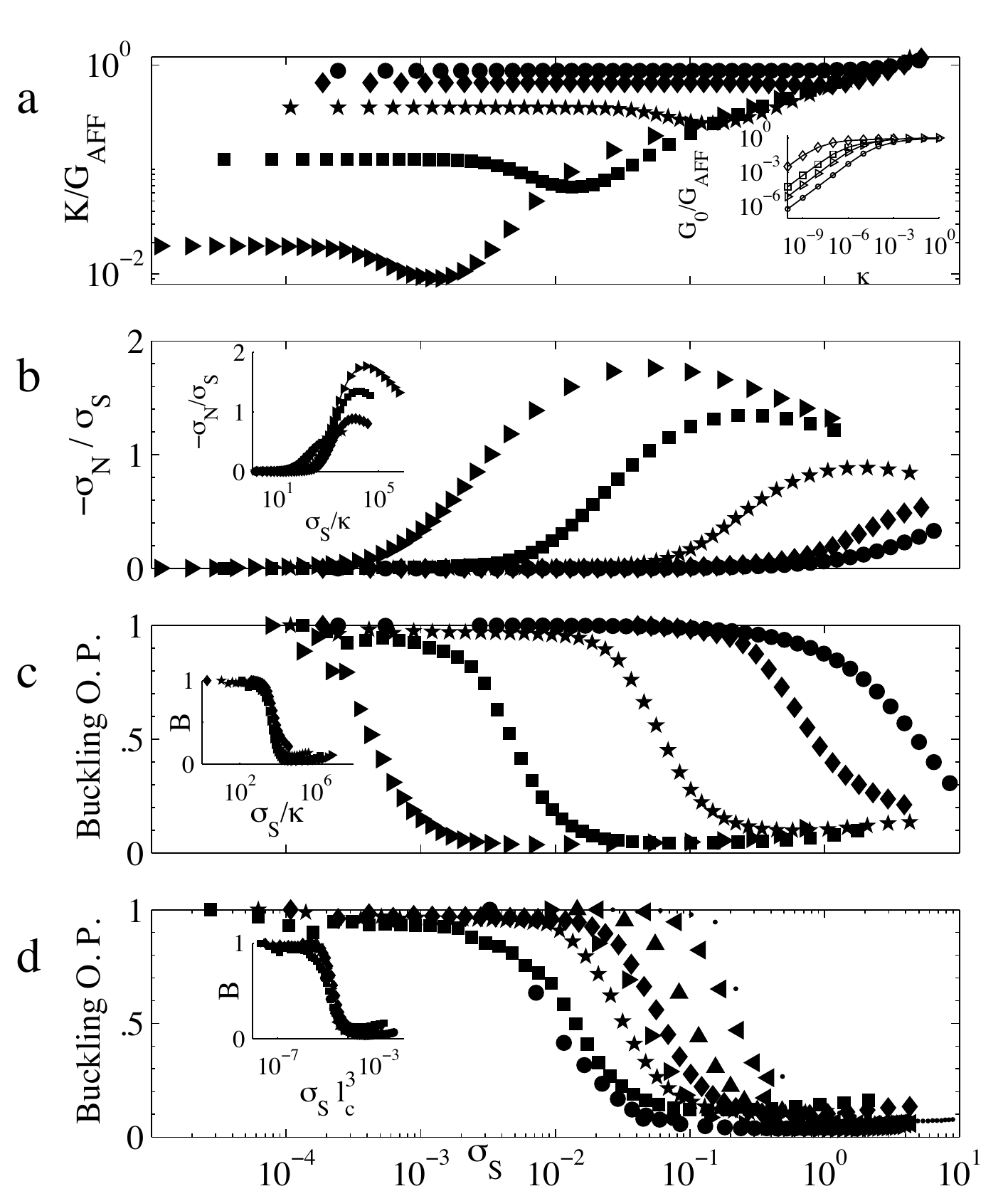}
\caption{In (a), (b) and (c), the differential shear modulus $K$ normalized by $G_{\mbox{\tiny{AFF}}}$, the ratio of normal to shear stress, and the buckling order parameter $B$ are all plotted versus shear stress for constant density $L/\ell_c=15$ and for various bending stiffnesses $\kappa=10^{-7}$ ($\blacktriangleright$), $\kappa=10^{-6}$ ($\blacksquare$), $\kappa=10^{-5}$ ($\bigstar$), $\kappa=10^{-4}$ ($\blacklozenge$), $\kappa=10^{-3}$ ($\bullet$).
In the inset of (a), the ratio $G_0/G_{\mbox{\tiny{AFF}}}$ is plotted versus the bending stiffness $\kappa$ for different densities $L/\ell_c=9$ ($\bullet$), $13$ ($\blacktriangleright$), $20$ ($\blacksquare$) and $40$ ($\blacklozenge$). Here, $G_0$ grows linearly with $\kappa$ for soft filaments and saturates to $G_{\mbox{\tiny{AFF}}}$ for stiff ones; $G_0/G_{\mbox{\tiny{AFF}}}$ also increases with density.
Insets in (b) and (c) show data collapse for $-\sigma_N/\sigma_S$ and $B$ when plotted against $\sigma_S/\kappa$.
In (d) $B$ is plotted for constant $\kappa=10^{-5}$ and various densities $L/\ell_c=11$ ($\bigstar$), $L/\ell_c=13$ ($\blacklozenge$), $L/\ell_c=15$ ($\bullet$), $L/\ell_c=20$ ($\blacktriangleright$), $L/\ell_c=30$ ($\blacktriangle$), $L/\ell_c=40$ ($\blacktriangleleft$), $L/\ell_c=50$ ($\blacktriangledown$) and $L/\ell_c=60$ ($\cdot$). The inset in (d) shows the collapse of these data when plotted against $\sigma_S\ell_c^3$. Together with the insets in (b) and (c), this shows that the onset of nonlinearity coincides with the buckling transition.
}
\label{ComparisonVsSigma}
\end{figure}

\begin{figure}
\centering
\includegraphics[width=250 pt]{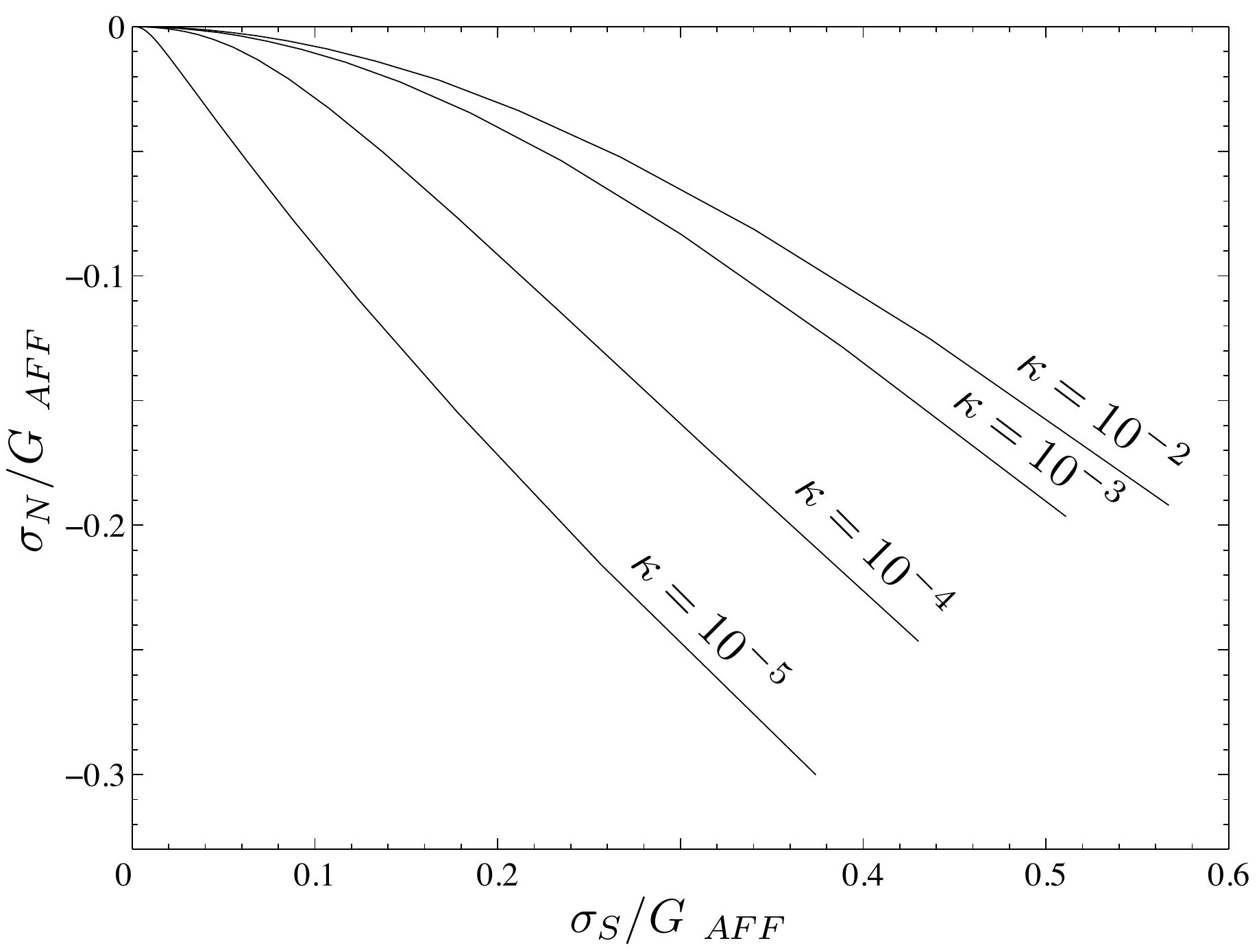}
\caption{Normal stress $\sigma_N$ versus shear stress $\sigma_S$ (both normalized by $G_{\mbox{\tiny{AFF}}}$) for density $L/\ell_c=13$ and various filament bending stiffness.
For stiff filaments, $\sigma_N\sim\gamma^2\sim\sigma_S^{\ 2}$ over a large initial stress region. For softer filaments this region becomes smaller so that on this linear scale it appears that $\sigma_N\sim\sigma_S$.} \label{NvsS}
\end{figure}

In Fig.~\ref{ComparisonVsSigma}a we plot the shear modulus $G=\sigma_{S}/\gamma$ and differential shear modulus $K=d\sigma_{S}/d\gamma$ normalized by the affine linear modulus $G_{\mbox{\tiny{AFF}}}=
\frac{\pi}{16}\frac{\mu}{L}\left(\frac{L}{\ell_c}+2\frac{\ell_c}{L}-3\right)$, which corresponds to purely compression/extension deformation without bending filaments \cite{Head03}.
This represents an upper bound for $G$ and $K$ at small strains, since bending modes can only lower the energy for given boundary conditions.
For these small strains, a linear regime with constant $K=G=G_0$ is seen in all cases, while nonlinearities such as stiffening and/or softening occur at higher strains.
The linear modulus $G_0$ increases both with density and bending stiffness in agreement with previous work \cite{Head03,Wilhelm03,Onck}.
Specifically, $G_0\propto \kappa$ for floppy filaments, while $G_0$ saturates to $G_{\mbox{\tiny{AFF}}}$ for stiff filaments (inset) or high densities.

\begin{figure}
\centering
\includegraphics[width=250 pt]{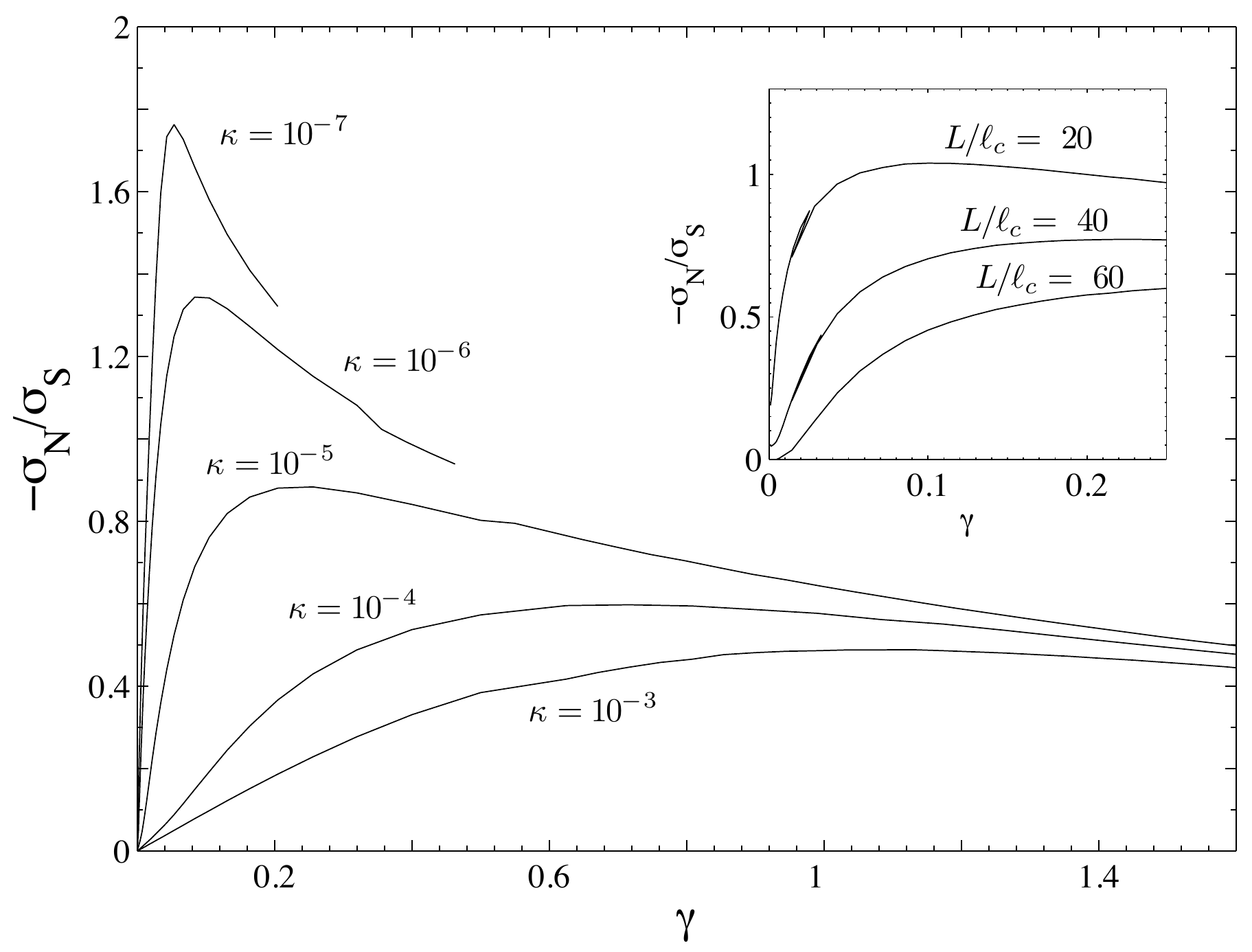}
\caption{The ratio of normal to shear stress versus applied strain $\gamma$ for constant density $L/\ell_c=15$ and various filament bending stiffnesses.
On decreasing $\kappa$ a peak grows, becoming more pronounced and moving to smaller strain.
For large strain the curves depend weakly on $\kappa$, showing a regime dominated by stretching only.
The inset shows the same ratio $-\sigma_N/\sigma_S$ for constant stiffness $\kappa=10^{-5}$ and different densities.
} \label{NoverS}
\end{figure}

In Fig.\ \ref{NvsS} we directly compare the shear and normal stresses, by plotting $\sigma_{N}$ vs $\sigma_{S}$, both of which have been normalized by $G_{\mbox{\tiny{AFF}}}$.
For stiffer filaments we see an approximate quadratic dependence, which is to be expected by symmetry: since $\sigma_{N}$ must be an even function of strain, it is generally expected to have an initial quadratic dependence for small strain, whereas $\sigma_{S}$ is expected to be approximately linear in strain.
Interestingly, for more flexible filaments, this quadratic regime shrinks and an approximately linear dependence ($-\sigma_{N}\simeq\sigma_{S}\sim\gamma$ is observed, in which the two stresses are of comparable magnitude.
This is consistent with what was found experimentally in Ref.\ \cite{Janmey07}.
This behavior can be understood in simple terms for highly asymmetric extensional response of filaments, in which both $\sigma_{S}$ and $\sigma_{N}$ are dominated by the stretched filaments oriented near $45^{\circ}$ to the strain direction\cite{Janmey07}.
Our results suggest that the crossover from quadratic $\sigma_{N}\sim\sigma_{S}^2$ behavior occurs for smaller strains with more flexible filaments, which can be tested experimentally.

To test this more directly, we plot the ratio $-\sigma_N/\sigma_S$ vs $\gamma$ in Fig.~\ref{NoverS}.
Here, we see that the small-strain regime, characterized by $\sigma_{N}\sim\gamma^2$ and $\sigma_N/\sigma_S\sim\gamma$, decreases with increasing flexibility of filaments or decreasing network concentration. Interestingly, this is opposite to the predicted behavior for thermally fluctuating networks in Ref.\ \cite{Janmey07}. Thus, the normal stresses may provide an experimental signature for thermal vs non-thermal systems.
We also observe an apparent convergence of all networks for increasing strain, to a regime in which the two stress components are comparable.
As the filaments become softer to bending, we observe a peak, with normal stresses exceeding shear stresses for small strains.
This may provide an explanation for the observation in Ref.\ \cite{Janmey07}, where the magnitude of normal stresses exceeded the shear stresses.

As argued in Ref.\ \cite{Janmey07}, one possible origin of negative normal stresses can be the asymmetric extensional response, or force-extension relation of the constituent filaments.
Even though the individual filaments in our model are assumed to be Hookean springs, with a symmetric force extension curve, the fact that they can bend and buckle effectively gives rise to a softer response to compression and an asymmetric force-extension relation\cite{FloppyModes}.
To test whether this can explain our results, we measure the relative importance of buckling with an order parameter $B$ defined by $\frac{dE_{\mbox{\tiny{C}}}}{d\gamma}/\frac{dE_{\mbox{\tiny{TOT}}}}{d\gamma}$.
Here, the numerator refers to the incremental change in the elastic compression energy of the filament segments under compression in the network for a small strain step $d\gamma$, evaluated as a function of the initial state of network strain $\gamma$.
The denominator refers to the incremental change in total elastic energy.
This is normalized by its initial, small-strain value, where no buckling is expected.
Thus, this order parameter should have initial value of unity, and should then decrease as filaments begin to buckle.
We plot this in Fig.\ \ref{ComparisonVsSigma}, along with both $K/G_{\mbox{\tiny{AFF}}}$ and $-\sigma_N/\sigma_S$ for several different networks of the same density, but with differing bending stiffness.
We see that the onset of significant buckling, indeed, coincides with the onset of nonlinearities in the rheology.
Furthermore, classical Euler buckling theory predicts that the threshold force for buckling is proportional to $\kappa$.
By plotting both $B$ and $-\sigma_N/\sigma_S$ versus $\sigma_S/\kappa$, we see collapse of the curves for various $\kappa$ (see insets), consistent with the dominant role of buckling in controlling the development of large normal stresses.
In Fig.\ \ref{ComparisonVsSigma}d we plot $B$ for various $L/\ell_c$ and we see that the curves collapse (inset) by plotting against $\sigma_S\ell_c^3$, confirming that the buckling order parameter actually accounts for the collective buckling of intercrosslink segments \cite{footnote}.

We have simulated networks of elastic rods with up to 200,000 degrees of freedom, and for a range of strains up to 2.
We find that coincident with the development of other nonlinear elastic properties, such networks very generally exhibit negative normal stresses of magnitude up to and even exceeding the corresponding shear stresses.
This highly unusual material property arises as a cooperative effect of filaments whose individual elastic behavior is linear.
Qualitatively, the normal stress behavior we observe can be understood, however, in terms of the nonlinear force extension that results from Euler buckling of the filaments under compression \cite{Janmey07,FloppyModes}.
We demonstrate this directly by measuring the degree of buckling in our networks.
This can explain, in part, the apparent generality of negative normal stresses reported in recent experiments on a number of different stiff biopolymer systems\cite{Janmey07}.
Furthermore, our results suggest a possible way to distinguish experimentally between alternative theories of thermal\cite{MacKPRL95,Gardel04,Storm05} vs athermal\cite{Onck} origins of nonlinear elasticity in biopolymer networks: we find that, the characteristic strain for the onset of negative normal stress increases with network concentration, in contrast with the prediction for thermal networks\cite{Janmey07}.

\begin{acknowledgments}
We thank C.\ Broedersz, M.\ Das, and P.\ Janmey for useful discussions.
\end{acknowledgments}


\begin{thebibliography}{prsty}
\bibitem{footnote}{For a rod of length $\ell_c$ the Euler buckling thershold force is $f_{\mbox{\tiny{crit.}}}\sim \kappa/\ell_c^2$ while the macroscopic stress is proportional to the force per filament $f_c$ times the flux of filaments crossing the surface which is proportional to $1/\ell_c$. Then we obtain the relation $\sigma_S^{\mbox{\tiny{crit.}}}\sim\kappa/\ell_c^3$.}
\end{thebibliography}
\end{document}